\documentstyle[sprocl,epsf]{article}
\def\be{\begin{equation}}
\def\ee{\end{equation}}
\def\bea{\begin{eqnarray}}
\def\eea{\end{eqnarray}}

\begin{document}
\title{ Saturation of Gluon Density at Small x\footnote{to appear in
the proceedings of DPF 96, August 10-15, Minneapolis, Minnesota.}}
\author{Jamal Jalilian-Marian }
\address{Physics Dept., Univ. of Minnesota,
Minneapolis, MN 55455}

\maketitle\abstracts{Computing parton distribution functions in QCD 
is a formidable task.
 It is intrinsically non-perturbative and and thus hard to calculate
from first principles. On the other hand, knowledge of these distribution
functions is required in any dynamical process involving hadrons. This 
note describes a new approach to the problem. It is shown how this new
approach leads to saturation of the gluon density at small $x$.}

\section{Introduction}
In perturbation theory, one considers evolution of soft gluon density
in $x$, $Q^2$ or both. In the double log approximation of DGLAP 
formalism, terms of the form $\alpha \ln (x) \ln (Q^2)$ are summed. This
is done by considering the so-called ladder diagrams where both
longitudenal and transverse momenta are ordered.

For very small values of $x$ it may be more appropirate to sum terms
of the form $\alpha \ln (x)$. This is done in BFKL formalism where one
has ordering in longitudenal momenta while assuming that everything
is happening at more or less the same transverse area. It is important to
realize that both DGLAP and BFKL are evolution equations and as such 
require initial parton distributions as input.

Regardless of whether BFKL or DGLAP is more appropirate for description
of soft gluons, they both have one outstanding feature in common; they
both predict a sharp rise of soft gluon density at very small values 
of $x$ and thus would eventually lead to violation of unitary bound 
on physical cross sections.

However, the above picture of harder partons spliting into softer
ones with no further interaction among them is perhaps a bit too naive.
Physically we know that at very small values of $x$, the parton density 
will be high and they will spatially overlap. The effects of parton 
recombination, screening etc., will therefore be important. This will
eventually lead to saturation of soft gluon density at small $x$.

It may be more natural to consider soft gluons not as quasi-free partons
but rather as classical fields with large amplitudes. After summing the 
effects of high density into the clasical fields one can do 
perturbation theory in the background of these large classical fields.

\section{McLerran-Venugopalan Model}
In the Mclerran-Venugopalan model of small $x$ gluons, one considers
a large nucleus traveling with very high velocities and thus
highly Lorentz contacted. It therefore has a high valence quark 
surface density. This approach can also be used for hadrons at 
sufficiently small values of $x$ where the parton density is large.
 The color charge density $\rho$ is the only dimentionfull
parameter in the problem. Therefore the strong coupling constant 
$\alpha _{s}$, as a function of $\rho$, will be small for large charge
densities so that one can use weak coupling methods. Since the coupling 
is weak, the valence quarks do not lose an appreciable fraction of their 
momenta and stay on straight line trajectories and are therefore 
static sources of color charge. This is the no-recoil approximation.
Also, due to the high density of glue, one can treat them as classical 
charges. This corresponds to taking a higher dimentional representation
of the color algebra. 
Treating the valence charges as classical leads to computing expectation
values with a Gaussian distribution of the form
\be
\int [d\rho]\,\, exp\big{\{}{{-1}\over {2\mu ^2}} \int d^2 x_t 
\rho ^2 (x_t)\big{\}}
\label{eq:GD}
\ee
where $\mu^2 $ is the average color charge squared per unit area.

Alternatively, one can solve the Yang-Mills equations in the presence of
this static color charge. Once a solution is found, one can then compute 
the distribution function with the above Gaussian weight. It was shown
in~\cite{MV} that this leads to a distribution function of the
 Weizs\"{a}cker-Williams
form for soft gluons. Quantum corrections to the classical results were also
computed and similar to standard perturbation theory, it was found that
there are potentially large logs of ratio of longitudenal momenta~\cite{QC}.
 This led to a longitudenal structure for the color charge density $\rho$. 
Then the problem to consider was to solve the classical equation of motion 
in the presence of this rapidity dependant charge density. 

Working in light-cone gauge $A^+=0$ and using light cone notation
$x^{\pm} \sim (t\pm z)$ and $y \sim \ln (x^-)$,
we have 
\be
D_i {{d}\over {dy}} A_i = g^2 \rho (y, x_t ).
\ee

One can write a formal solution to this equation in terms of $\rho$ and
then compute 
the corelation function $G^{aa}_{ii}\sim < A^{a}_{i} A^{a}_{i} >_{\rho} $
 where now both $\rho $ and $ \mu ^2$ depend on $y$. The result is 
\be
G^{aa}_{ii}= {{4(N^2_c - 1)} \over {N_c x^2_t}} 
\big[ 1- (x^2_t \Lambda ^2_{QCD})^{{{g^4 N_c }\over {8\pi}} \chi (y, Q^2)
 x^2_t } \big]
\ee
where 
\be
\chi(y,Q^2)= \int_{y}^{\infty} dy^{\prime} \mu^2 (y^{\prime} ,Q^2)
\nonumber
\ee
is the average color charge squared per unit area at 
rapidity $y$ and resolution scale $Q^2$. 

\section{RG Equation for $\chi$}
Next we need to determine $\chi(y, Q^2)$. We start with the valence charge 
density and then include hard gluons into the valence charge step by step. 
In other words, we integrate out higher x gluons perturbatively. This 
generates an effective Lagrangian and  renormalizes the charge
density. Iterating this procedure one can derive a RG equation for $\chi$. 
 Diagramatically this can be represented as 

\hskip 2cm
\epsfxsize=8.0cm
\epsffile{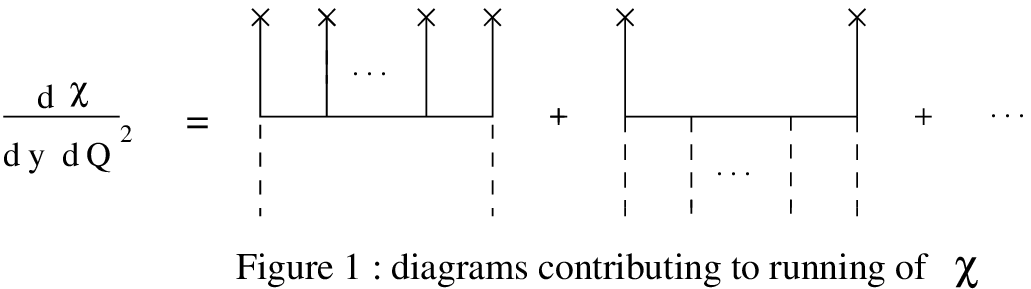}
\label{fig:fig1}

In the first diagram, the propagator of the hard field is in the background
of classical fields. This propagator was computed in~\cite{PR}. In the second 
diagram the hard field propagates in the background of soft modes. This is
where Sudakov and non-Sudakov form factors are expected to show up. There 
are also diagrams where the hard and soft modes are mixed. 

As a first approxiamation, we will use the following ladder diagram on the
right hand side of our RG equation where the vertices are eikonalized.
It is know that $\ln (x)$ terms come from eikonal vertices.

\hskip 4cm
\epsfxsize=8.0cm
\epsffile{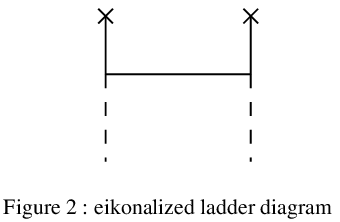}
\label{fig:fig2}

Solid and dashed lines are hard and soft gluons respectively. With this 
approximation, the RG equation becomes 
\be
{{d^2 \chi }\over {dy dQ^2 }} \sim G^{aa}_{ii}(y,Q^2).
\ee
It should be emphasized that the right hand side is still a non-linear 
functional of $\chi(y,Q^2) $.

\section{Solution to RG Equation}
In the high $Q^2$ region ($Q^2 \gg \alpha ^2 \chi $), it can be shown that
our RG equation reduces to the standard DGLAP evolution equation.
Assuming $\alpha $ to be independent of $Q^2$, the approximate solution is 
\be
\chi \sim exp \big( 2 \sqrt{{{N_c \alpha_s}\over {\pi}} y \ln Q^2 }\big).
\nonumber
\ee
In the intermediate $Q^2$ region ($Q^2 \sim \alpha^2 \chi $), the effects of
the background fields as well as the soft fields are expected to be important
quantitatively. However, here we are interested in the qualitative behavior
of our RGE. Assuming that, as in BFKL, the transverse phase space is a 
constant, we get the following BFKL like behavior
\be
\chi \sim exp \big(\# {{N_c \alpha_s }\over {\pi}}y\big).
\nonumber
\ee

Finally in the low $Q^2$ region ($Q^2 \ll \alpha^2 \chi $), our RG equation
saturates and the solution is of the form
\be
\chi = \chi_\circ + \kappa (y_\circ - y) Q^2
\nonumber
\ee
where $\kappa $ is a slowly varying function of $Q^2$. This leads to 
saturation of the gluon distribution function at small $x$. As a result, 
cross sections computed with this distribution function would satisfy
unitarity bounds. To illustrate this saturation, let us consider the 
behavior of $\chi(y,Q^2)$ at some fixed $Q^2$ as $x$ decreases. In the high 
transverse momentum region ($k^2_t \gg \alpha^2 \chi$), the distribution 
function will grow as some power
of $x$. As $x$ gets smaller, $\alpha^2 \chi$ will eventually become larger
than $Q^2$ and we  will be in the saturation region (the region between 
$\Lambda^2_{QCD}$ and $\alpha^2 \chi$) where the distribution function is a 
slowly varying function of $k_t^2$. This is shown below.

\hskip 1cm
\epsfxsize=8.0cm
\epsffile{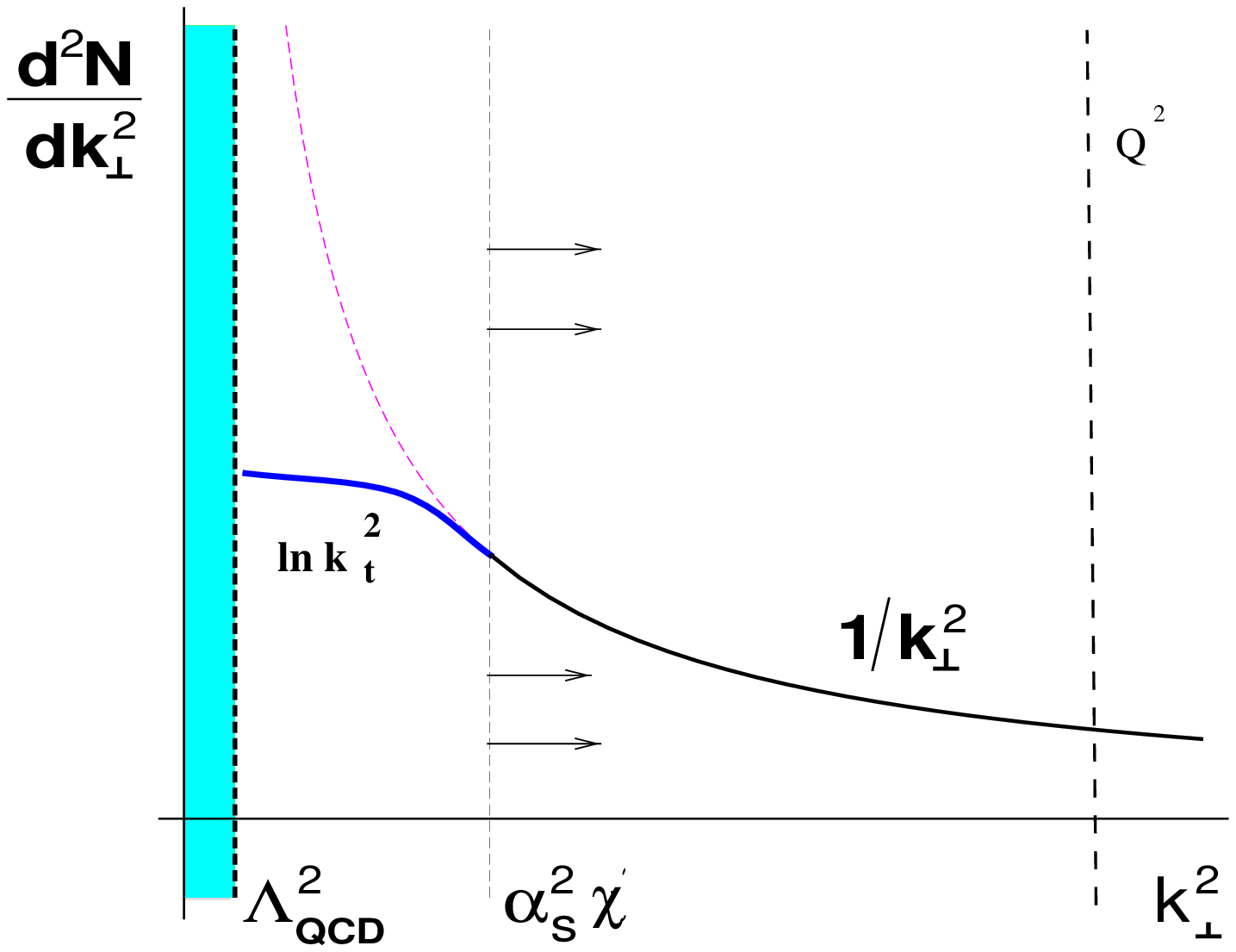}
\label{fig:fig3}

Notice that this saturation is due to shrinking of the transverse phase 
space as indicated above.

\section{Acknowledgments} The work presented here has been done in 
collaboration with Alex Kovner, Larry McLerran and Heribert Weigert.
For a full derivation of the results presented here see~\cite{JJM}.


\begin{thebibliography} {99}

\bibitem{MV}  L. McLerran and R. Venugopalan,{\it Phys. Rev.}\,{\bf D49},
 2233 (1994);
\,{\bf D49}, 3352(1994);\, {\bf D50}, 2225(1995).

\bibitem {QC} A. Ayala, J. Jalilian-Marian, L. McLerran and 
R. Venugopalan, {\it Phys. Rev.}\,{\bf D53}, 458 (1996).

\bibitem {PR}  A. Ayala, J. Jalilian-Marian, L. McLerran and 
R. Venugopalan, {\it Phys. Rev.}\,{\bf D52}, 2935 (1995).

\bibitem {JJM} J. Jalilian-Marian, A. Kovner, L. McLerran and
H. Weigert, hep~-~ph/9606337. 

\end{thebibliography}
\end{document}